\documentclass[aps,twocolumn,pra,twoside,amssymb,amsmath]{revtex4}
\usepackage{amssymb}
\usepackage{graphicx}
\usepackage{amsmath}
\usepackage{colordvi}
\usepackage{bbm}
\usepackage{verbatim}
\usepackage{float}
\usepackage{dcolumn}
\usepackage[colorlinks,linkcolor=blue,anchorcolor=blue,citecolor=blue,urlcolor=blue]{hyperref}
\usepackage{mathrsfs}

\begin{document}
\title{Chiral interaction enhanced magnon bundle emission}
\author{Zhicai Chen$^{1}$, Deyi Kong$^{2}$, Chengdeng Gou$^{3}$,  Xiangming Hu$^{1,}$\footnote{xmhu@ccnu.edu.cn} and Fei Wang$^{2,}$\footnote{feiwang@hbut.edu.cn}}
\affiliation{
  $^{1}$\mbox{College of Physical Science and Technology, Central China Normal University, Wuhan 430079, China} \\
  $^{2}$\mbox{School of Science, Hubei University of Technology, Wuhan 430068, China}
  $^{3}$\mbox{School of Mathematics and Physics, Qinghai University, Xining 810016, China}}
\begin{abstract}
In this paper, we suggest a chiral interaction scheme to enhance magnon bundle emission by placing a qubit and a magnon into a cascaded-cavity setup, respectively. It is found that the unidirectional interaction prolongs the lifetime of the target excited state, thereby suppressing the magnon re-excitation and promoting both the average purity and number of two-magnon bundles. Consequently, the chiral interaction not only offers directional control but also improves the quality of the multi-magnon source, which may find potential applications in quantum information processing. 
\end{abstract}

\maketitle

\section{Introduction}
Chiral light-matter interactions, which are closely related to the photonic spin-momentum locking and time-reversal symmetry breaking, in essence, arise when the coupling between a quantum emitter and light depends on the propagation direction of light and the polarization of the transition dipole moment of the emitter \cite{lodahl2017chiral}. Due to this unique physical effect, chiral interaction provides a promising platform to realize the nonreciprocal devices including  optical isolators \cite{sayrin2015nanophotonic,dong2021all} and circulators \cite{scheucher2016quantum,xia2018cavity}, and has wide potential applications in  quantum information processing \cite{sollner2015deterministic,suarez2025chiral}, quantum routers for realizing large-scale quantum networks \cite{yan2022realization, palaiodimopoulos2024chiral,liu2025nonreciprocal} and quantum light source \cite{Huang2022Exceptional,lu2025chiral}. Recently, chiral light-matter interactions are no longer confined to traditional atomic and quantum-dot-based platforms \cite{suarez2025chiral}. Instead, solid-state systems such as microcavity polaritons \cite{chen2024chiral2d,wang2023spin} and two-dimensional layered materials \cite{guddala2021all}, when integrated into waveguides, cavities, and ring resonators, now provide versatile settings for investigating chiral interactions in quantum optics.

In parallel, bundle emission has received continuous interest in multi-quantum physics. It refers to a situation where multiple correlated quanta are consecutively released from a quantum emitter, in which the quanta within the same bundle are strongly correlated while the neighboring bundles are antibunched. Such quantum light sources have important applications in quantum technology including multiplexed quantum communication \cite{llewellyn2020chip,meyer2022scalable}, entanglement generation \cite{bin2024entangled,bin2024nonreciprocal}, quantum metrology \cite{dowling2008quantum,joo2011quantum,qin2023unconditional} and quantum biology \cite{horton2013vivo,li2023two}. Initially, the typical scheme for generating $N$-photon bundle emission was proposed using Jaynes-Cummings model in cavity quantum electrodynamics (QED) \cite{munoz2014emitters}. Subsequently, a large number of theoretical proposals have been investigated extensively to prepare various types of bundle emission, including multi-level atomic systems \cite{chang2016deterministic,gou2022antibunched,gou2024antibunched}, acoustic cavity QED via Stokes process \cite{bin2020n}, waveguide-QED \cite{xing2024deterministic}, optomechanical systems \cite{zou2022dynamical}, magnon bundle \cite{yuan2023magnon,zhao2025heralded}, and the hybrid bundle \cite{bin2024entangled,bin2024nonreciprocal,gou2024hybrid,wang2025photon}.

In general, in $N$-photon bundle case, the cavity field would be indirectly driven into a high Fock state by an external field, $|0\rangle\xrightarrow{\textrm{drive}}|N\rangle$, then probabilistically emits $N$ photons from the superposition to the vacuum state, $|N\rangle\xrightarrow{\textrm{decay}}|0\rangle$. At the end, the atom decays back to its ground state with a slower decay rate compared with the cavity dissipation, providing the opportunity to separate the successive emitted bundles. Obviously, the quality of the bundle is essentially determined by the following two factors: the emission number (brightness) and the purity. Accordingly, it is useful to enhance the emission number by strengthening the atom-cavity coupling via parametric amplification technology \cite{chen2025enhanced}. On the other hand, the purity of bundle emission would be improved when the two-photon dissipative processes dominate over single-photon losses\cite{xiong2025two}. Moreover, for a parity-protected system, it has been shown that the even (2n)-photon bundles are preferred while the odd-photon emission is nearly blockaded, leading to higher purity of two-photon bundle \cite{bin2021parity}. Interestingly, we seek a completely different method to promote the brightness and the purity of magnon bundle simultaneously by utilizing chiral interaction. The magnon is regarded as the quanta of collective spin excitations in magnetic crystals such as yttrium iron garnet (YIG), possessing unique properties of high spin density, extremely low magnetic damping \cite{zhang2014strongly,tabuchi2014hybridizing,bourhill2016ultrahigh,kostylev2016superstrong}, and strong compatibility with microwave photon, optical photon, phonon, NV-center, superconducting qubits \cite{yuan2022quantum,huebl2013high,osada2016cavity,zhang2016optomagnonic,zhang2016magnomechanics,tabuchi2015coherent,lachance2017resolving,lachance2020entanglement}. These features make magnonic systems a promising magnetic platform for information exchange and storage \cite{kimble2008quantum,dong2015optomechanical,Li2021QuantumNetwork,Lu2025Optomagnonic}. Up to now, the magnon bundle emission has attracted interest in a strongly dissipative magnet \cite{yuan2023magnon}, hybrid magnon-cavity system \cite{bin2024entangled,bin2024nonreciprocal}, and hybrid ferromagnet-superconductor system \cite{gou2024hybrid}.

\begin{figure}[t]
  \centering
	\includegraphics[scale=0.43]{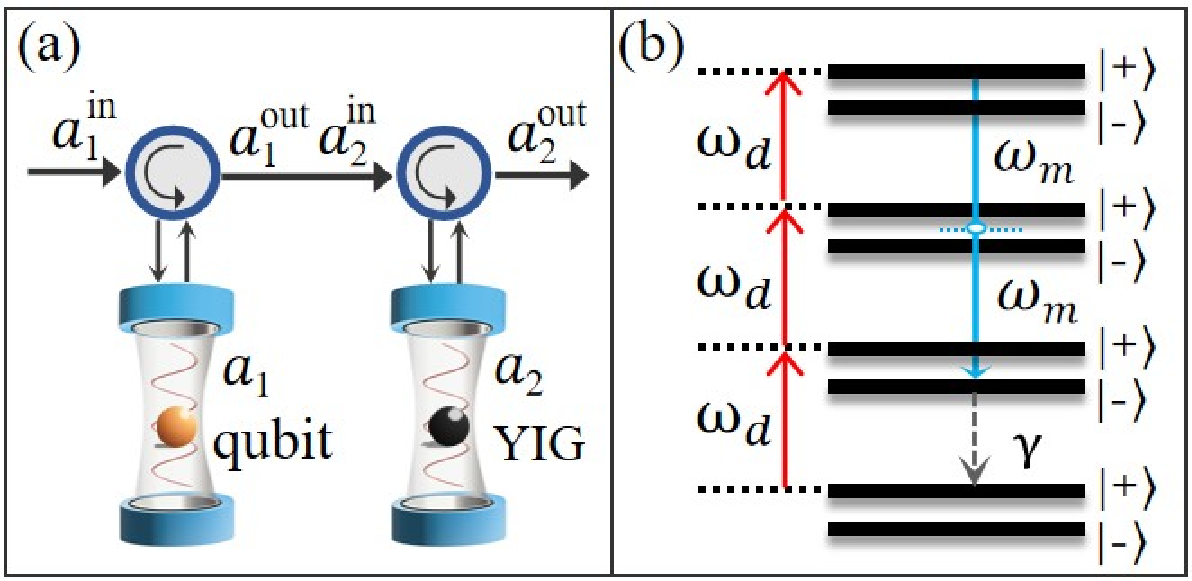}
    \caption{(a) The system schematic involves cascading the output of one cavity into another, where a qubit driven by a strong microwave field is placed in the left cavity and a YIG sphere is placed in the right cavity. Experimentally, the circulators are required to enforce unidirectional propagation of fields. (b) Schematic of the mechanism for two-magnon bundle emission. \label{Fig1}}
\end{figure}

In this paper, by placing a qubit and a YIG sphere into two cascaded cavities \cite{gardiner1993driving,carmichael1993quantum,tan2021einstein,combes2017slh,Stannigel2012Driven}, respectively, the output field from the first cavity is input into the second cavity continuously. With this typical cascade two-cavity system, the chiral light-matter interaction is established  by an engineered dissipative process and the non-Hermitian interaction Hamiltonian is derived by adiabatically eliminating the cavity fields, corresponding to the absence of a process where the magnon is annihilated while the qubit is flipped from the ground state to the excited state. It helps prolong the lifetime of the target state, thus suppressing the re-excitation of magnons. Therefore, the quality of the magnon bundles can be enhanced due to the unidirectional coupling between the atom and the magnon. Our theoretical results demonstrate that both the average bundle purity and the average bundle number are increased compared with the normal case of symmetric coupling.

This paper is organized as follows: Section \ref{sec2} introduces the proposed model and equations of our scheme. In Section \ref{sec3}, through comparison with symmetric coupling, we demonstrate a comprehensive analysis of chiral coupling enhanced magnon bundle emission. Conclusions are summarized in Sec. \ref{sec4}.

\section{model and equations}\label{sec2}
The system schematic is shown in Fig. \ref{Fig1}(a), where the setup consists of two cascaded cavities. A qubit driven by a strong microwave field is placed in the left cavity, while a YIG sphere is placed in the right cavity. The output of the left cavity serves as the input to the right cavity, but not vice versa. Through this cascaded cavity setup, unidirectional coupling between the qubit and magnon is established, due to an engineered dissipative process. According to Ref. \cite{gardiner1993driving}, 
the master equation for our cascaded system can be written as follows in an appropriate rotating frame $(\hbar=1)$
\begin{eqnarray}  \label{master}
\nonumber
    \frac{\partial\rho}{\partial t} =&-&i[H_{\textrm{aq}}+H_{\textrm{am}},\rho] \\ \nonumber
    &+&\sum_{j=1}^{2} \frac{\kappa_j}{2}(2a_j \rho a_j^\dagger-a_j^\dagger a_j\rho-\rho a_j^\dagger a_j) \\  \nonumber
    &+&\frac{\kappa_{m}}{2}(2m\rho m^\dag-m^\dag m\rho -\rho m^\dag m ) \\  \nonumber
    &+&\frac{\gamma_{q}}{2}(2\sigma_{12}\rho\sigma_{21}-\sigma_{22}\rho-\rho\sigma_{22}) \\  
    &-&\sqrt{\kappa_1\kappa_2}([a_2 ^\dagger,a_1\rho]+[\rho a_1^\dagger,a_2]).
\end{eqnarray}
Here $\gamma_q$, $\kappa_{m}$ and $\kappa_{1,2}$ are the dissipation rate of qubit, magnon and cavities.
The operators $a_1,a_2$ ($a_1^\dag ,a_2^\dag$) and $m$ ($m^\dagger$) are the annihilation (creation) operators for the cavity fields and magnon mode, respectively. Additionally, $\sigma_{ij} = |i\rangle\langle j|$ (for $i,j=1,2$) are the projection operators when $i=j$ and the spin-flip operators when $i\neq j$. The Hamiltonians $H_{\textrm{aq}}$ and $H_{\textrm{am}}$ read  

\begin{figure}[htbp]
  \centering
	\includegraphics[scale=0.55]{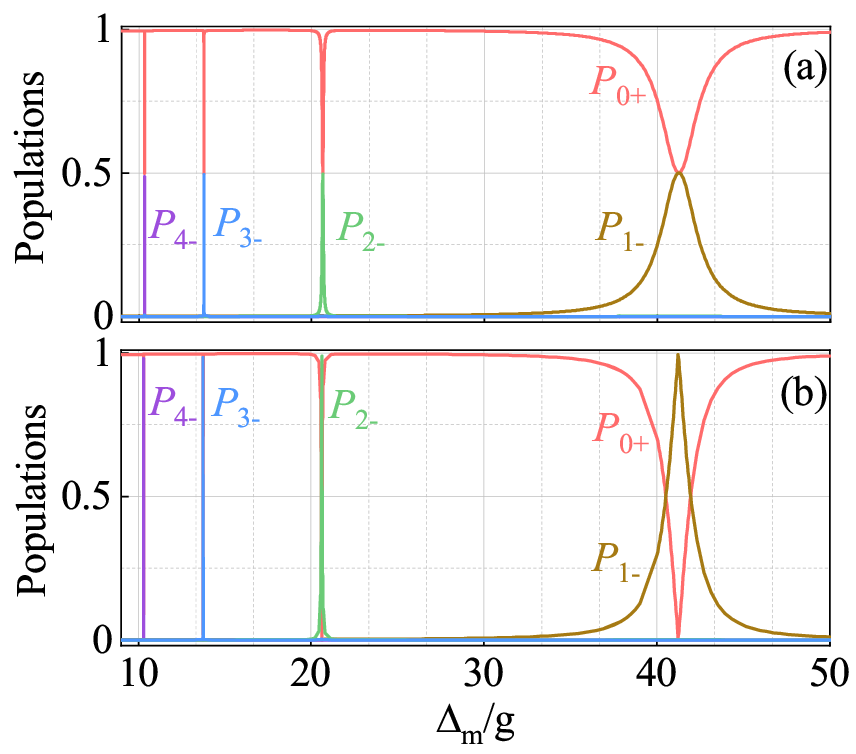}
    \caption{Time-averaged population of states $\{|0,+\rangle, |1,-\rangle,\\ |2,-\rangle,...\}$ as a function of $\Delta_{m}$ without dissipation. The parameters are $\Delta_q = 10 g$, $\Omega=20g$ and different coupling strength (a) $g_1=g_2=g$, (b) $g_1=0,\,g_2=-ig$. \label{Fig2}}
\end{figure}

\begin{eqnarray}
\nonumber
H_{\textrm{aq}}&=&\Delta_{1}a_1^{\dagger}a_1+\Delta_q' \sigma_{22}+\Omega(\sigma_{21}+\sigma_{12})\\ 
&+&g_{aq}(a_1^\dagger\sigma_{12}+a_1\sigma_{21}), \\ \nonumber
H_{\textrm{am}}&=&\Delta_2a_2^{\dagger}a_2+\Delta_m' m^\dagger m + g_{am}(a_2^\dagger m + a_2 m^\dagger),
\end{eqnarray}
where $\Delta_{1,2}=\omega_{a_{1,2}}-\omega_d$, $\Delta_q'=\omega_{q}-\omega_d$, and $\Delta_m'=\omega_{m}-\omega_d$ denote the detunings of cavities, atom, and magnon from the driving field frequency, $\Omega$ is the Rabi frequency quantifying the coupling strength between the qubit and microwave, $\omega_d$ is the frequency of microwave field and $g_{\textrm{aq}}$ ($g_{\textrm{am}}$) is the coupling strength between the left (right) cavity and the qubit (magnon). The Hamiltonian $H_{\textrm{aq}}$ describes the interaction between qubit and left cavity, while $H_{\textrm{am}}$ indicates the interaction between magnon and right cavity.  

In the dispersive limit $(\Delta_1,\Delta_2\gg\kappa_1,\kappa_2)$, the cavity modes can be adiabatically eliminated and a more concise Hamiltonian is obtained. Following the standard technique, we derive the quantum Langevin equations (QLEs) of two cavities as follows:
\begin{eqnarray}
\nonumber
    \dot{a}_1= &-&({\kappa_1/2+i\Delta_1})a_1 -ig_{aq}\sigma_{12}, \\ 
    \dot{a}_2= &-&({\kappa_2/2+i\Delta_2})a_2 -ig_{am}m. 
\end{eqnarray}

\begin{figure*}[htbp]
  \centering
	\includegraphics[scale=0.52]{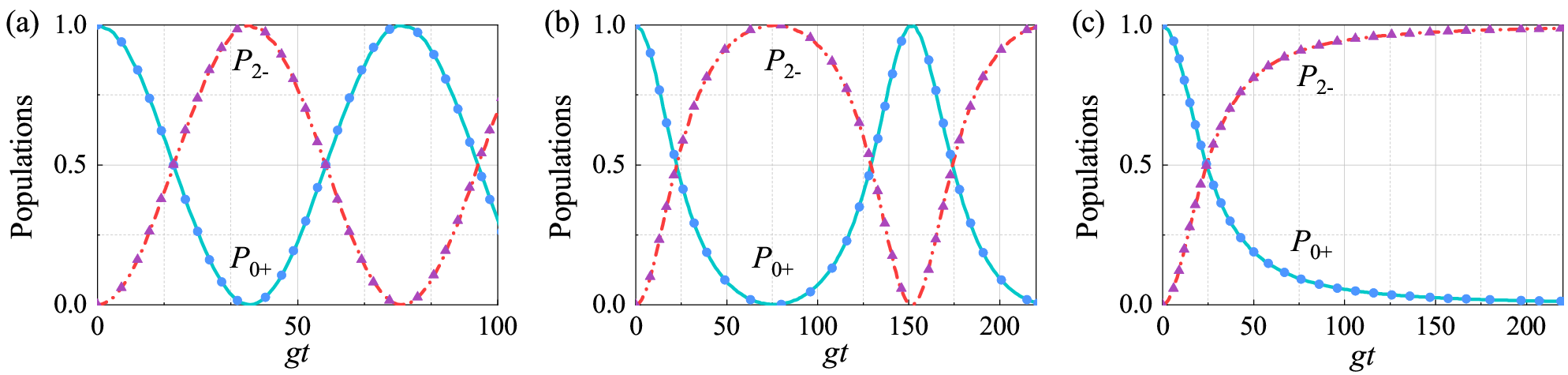}
    \caption{Time evolution of populations of states $|0,+\rangle$ and $|2,-\rangle$ at $\Delta_{m}\approx\tilde{\Omega}/2$ in the absence of system dissipation. The dashed and solid lines represent numerical results based on Hamiltonian of Eqs. (\ref{EqH_t}) while the triangular and circular points denote the analytical results based on Eq. (\ref{Eq10}). Different coupling strengths are used for (a) $g_1=g$, (b) $g_1=0.5g$ and (c) $g_1=0$. The other parameters are the same as those in Fig. \ref{Fig2}. \label{Fig3}}
\end{figure*}
Setting the left side of the above equations to zero and discarding the rapid evolution effect, we therefore deduce that
\begin{eqnarray} \label{trace} \nonumber
a_1=&-&\frac{ig_{aq}}{\kappa_1/2+i\Delta_1}\sigma_{12}, \\ 
a_2=&-&\frac{ig_{am}}{\kappa_2/2+i\Delta_2}m. 
\end{eqnarray}
Substituting the Eqs. (\ref{trace}) into the master equation of Eq. (\ref{master}), we can obtain 
\begin{eqnarray} \label{Eqnon}
    \nonumber
    \frac{\partial\rho}{\partial t} =&-&i\left(H_{\textrm{non}}\rho-\rho H_{\textrm{non}}^\dagger\right) \\ \nonumber
    &+&\frac{\kappa_{m}}{2}(2m\rho m^\dag-m^\dag m\rho -\rho m^\dag m ) \\  \nonumber
    &+&\frac{\gamma_{q}}{2}(2\sigma_{12}\rho\sigma_{21}-\sigma_{22}\rho-\rho\sigma_{22}) \\  
    &+&g(\sigma_{12}\rho m^\dagger+m\rho\sigma_{21}).
\end{eqnarray}

 Here we construct a non-Hermitian Hamiltonian to better elucidate the internal mechanism of our system. $H_{\textrm{non}}$ is written as
\begin{eqnarray}\label{Eqone} \nonumber
H_{\textrm{non}}&=&{\Delta}_m m^\dagger m + {\Delta}_q\sigma_{22}-ig\sigma_{12} m^\dagger \\ &+&\Omega(\sigma_{12}+\sigma_{21}),
\end{eqnarray}
where ${\Delta}_q\approx\Delta_q'-g_{aq}^2/\Delta_1$, ${\Delta}_m\approx\Delta_m'-g_{am}^2/\Delta_2$ and $g \approx g_{am}g_{aq}\sqrt{\kappa_{1}\kappa_{2}}/(\Delta_1\Delta_2)$ due to $\Delta_{1,2}\gg\kappa_{1,2}$.  The above non-Hermitian Hamiltonian explicitly demonstrates the unidirectional coupling between the magnon and the atom. Furthermore, we should point that cavity photon losses in the transmission channel, distinct from $\kappa_{1,2}$ , inevitably introduce additional dissipation for both the atom and magnon. Nevertheless, this additional dissipation can be suppressed by effectively optimizing the experimental parameters.
In the following section, we highlight the enhancement of magnon bundle emission by the one-way interaction through a comparison with the symmetric coupling case. Without loss of generality, we consider a Hamiltonian:
\begin{align} \label{EqH_t}
H &= \Delta_m m^\dagger m + \Delta_q\sigma_{22} + \Omega(\sigma_{21} + \sigma_{12}) \nonumber\\
  &  + g_1\sigma_{21}m + g_2\sigma_{12}m^\dagger.
\end{align}

When $g_1=0$, $ g_2=-ig$, the Hamiltonian $H$ describes the unidirectional coupling case and equals Eq. \ref{Eqone}; when $g_1=g_2=g$ (with $g$ real), it reduces to the conventional symmetric coupling case.

In this context, we adopt the strong driving conditions in the Mollow regime ($\Omega\gg g_1,g_2$) to generate the magnon bundle emission. Therefore, it is convenient to elucidate the mechanism of magnon bundle emission in the dressed-state picture.  The Hamiltonian of the strongly driven atom can be diagonalized as $\tilde{H}_\textrm{q}=\lambda_+\sigma_{++}+\lambda_-\sigma_{--}$, where $\lambda_{\pm}=\frac{1}{2}\left(\Delta_q\pm\tilde{\Omega}\right)$, and $\tilde{\Omega}=\sqrt{\Delta_q^2+4\Omega^2}$. Then, the dressed states can be expressed in terms of the bare states as follows:
\begin{eqnarray}\label{Eq_dress}
\nonumber
|-\rangle=\cos\theta|1\rangle -\sin\theta|2\rangle, \\
|+\rangle=\sin\theta|1\rangle +\cos\theta|2\rangle,
\end{eqnarray}
where $\sin\theta=\sqrt{\frac{1}{2}-\frac{d}{2\sqrt{1+d^2}}}$ and $\cos\theta=\sqrt{\frac{1}{2}+\frac{d}{2\sqrt{1+d^2}}}$ with $d={\Delta_q}/{(2\Omega)}$. In the dressed picture, the effective Hamiltonian of two-magnon resonance has been derived to understand the underlying mechanism of the bundle emission using adiabatic elimination approach in the Appendix. \ref{secA}. Similarly, for $N$-magnon resonance, the Baker-Campbell-Hausdorff formula is expanded to the $N$-th order and the corresponding effective Hamiltonian is derived under the rotating frame $\Delta_m\approx( \pm\tilde{\Omega}/N,0)$. The $N$-magnon resonance transition channels and corresponding frequency conditions are given as
\begin{figure*}[htbp]
  \centering
	\includegraphics[scale=0.6]{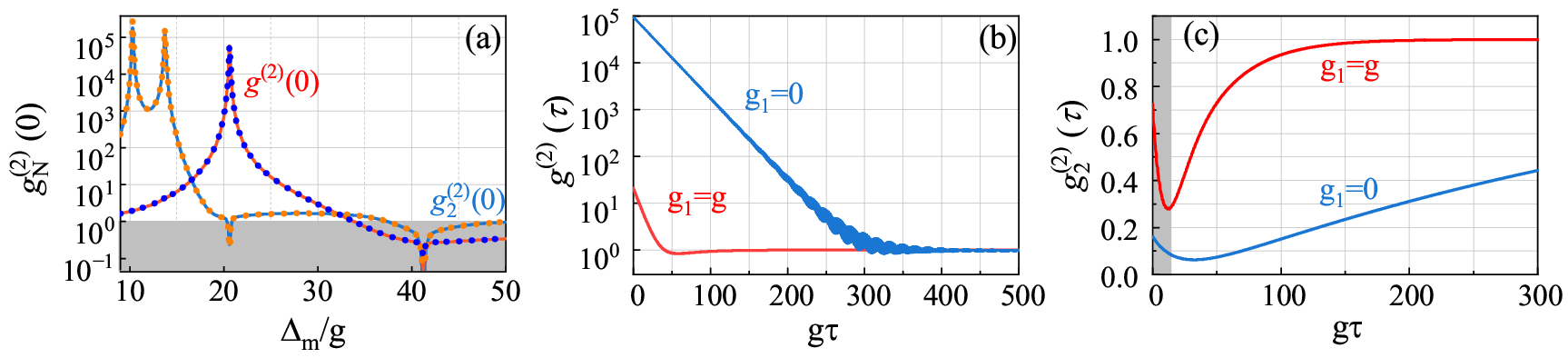}
    \caption{ (a) Equal-time second-order correlation functions $g^{(2)}_N(0)$ of one-way case versus the normalized detuning $\Delta_{m}/g$. Time-delayed second-order correlation functions (b) $g^{(2)}(\tau)$ and (c) $g_2^{(2)}(\tau)$ under the condition of two-magnon resonance. The gray shaded regions in (a) indicate $g_2^{(2)}(0)<1$, whereas that in (c) marks the invalid parameter interval $\tau<\tau_\textrm{min}$, where the bundle cannot be defined. In (a), the solid curves are based on the master equation (\ref{Eqnon}), while the dotted curves are obtained from the non-Hermitian Hamiltonian (\ref{Eqone}) together with the respective atomic and magnon dissipations. The parameters are $\Omega=20g$, $\Delta_q=10g$, $\kappa_m=0.1g$, and $\gamma_q=0.01g$.}  \label{Fig4}
\end{figure*}

\begin{eqnarray}\label{Eq8}
\nonumber
|+\rangle\xrightarrow{N}|-\rangle &:& \Delta_m\approx\frac{\tilde{\Omega}}{N}, \\ \nonumber
|-\rangle\xrightarrow{N}|+\rangle &:& \Delta_m\approx-\frac{\tilde{\Omega}}{N}, \\ 
|\pm\rangle\xrightarrow{N}|\pm\rangle &:& \Delta_m\approx0,
\end{eqnarray}
where $|i\rangle\xrightarrow{N}|j\rangle$ denotes the transition from the dressed-state $|i\rangle$ to  $|j\rangle$ while the magnon mode is in a high Fock state. In Fig. \ref{Fig1}(b), the mechanism for two-magnon bundle is illustrated.
Note that when $\Delta_m\approx0$, the retained effective Hamiltonian takes the form $\sum_{\ell=1}^{\infty}\left(C_{\ell,+}m^\ell\sigma_{++}+C_{\ell,-}m^\ell\sigma_{--}+\textrm{H.c.}\right)$. Under this resonance condition, all diagonal multi-magnon processes become resonant simultaneously, so a specific $N$-magnon bundle cannot be selectively realized. To verify theoretical calculations, we plot the time-averaged populations for states $\langle P_i\rangle$ as a function of $\Delta_{m}$ in Fig. \ref{Fig2}. Interestingly, it is worthwhile to notice that the populations transfer is different for the Hermitian and non-Hermitian cases. As shown in Fig. \ref{Fig2}(a), four resonance peaks appear with half of population exchange for $\langle P_{0+}\rangle=\langle P_{N-}\rangle=1/2$, in which the populations would oscillate back and forth between ground state $|0,+\rangle$ and $|N,-\rangle$. The resonance frequencies are in exact accordance with the theoretical results in Eq. (\ref{Eq8}). Differently, in Fig. \ref{Fig2}(b), we see that the positions of the peaks keep unchanged, at which the system would remain indefinitely in the target excited state $|N,-\rangle$. This is attributed to the fact that this term $g_1\sigma_{21}m$ is absent in the unidirectional coupling case. As a result, it gives rise to a situation where the lifetime of the state $|N,-\rangle$ is prolonged, which is beneficial to magnon bundle emission. 

To gain deeper insight into the physical mechanism of bundle emission, we derive the effective Hamiltonian at the resonance condition $\Delta_{m}\approx\tilde{\Omega}/2$ ( see the details in the Appendix \ref{secA}),
\begin{eqnarray} \label{Eq10} \nonumber
H_{\textrm{eff}}^1&=&g_{\textrm{eff}}^1m^2\sigma_{+-}+g_{\textrm{eff}}^2m^{\dagger 2}\sigma_{-+} \\ \nonumber
&=& \sum_{n=0}^{\infty}\left[\tilde{g}^1_{\textrm{eff}} |n,+\rangle\langle n+2,-|+\tilde{g}^2_{\textrm{eff}}|n+2,-\rangle\langle n,+|\right], \\
\end{eqnarray}
where $\tilde{g}^i_{\textrm{eff}}=g^i_{\textrm{eff}}\sqrt{(n+1)(n+2)},i=1,2$. It implies that the transition $|n,+\rangle\leftrightarrow|n+2,-\rangle$ is allowed while the other pathways that drive the system to non-target states are forbidden. When $g_1=g$, the system oscillates between states $|0,+\rangle$ and $|2,-\rangle$. However, for the case of $g_1=0$, the system is prepared from initial state $|0,+\rangle$ to target state $|2,-\rangle$ and then is trapped in this target state with 100$\%$ probability due to unidirectional interaction. Fig. \ref{Fig3} shows the time evolution of populations of states $|0,+\rangle$ and $|2,-\rangle$ by choosing different coupling strengths of $g_1$. In the symmetrical coupling case ($g_1=g$) in Fig. \ref{Fig3}(a), the so-called super-Rabi oscillation is demonstrated, corresponding to the condition of $\langle P_{0+}\rangle=\langle P_{2-}\rangle=1/2$ in Fig. \ref{Fig2}(a) at $\Delta_{m}\approx 20.6g$. By choosing $g_1=0.5g$, as shown in Fig. \ref{Fig3}(b), we see that the system prefers to stay in the target state $|2,-\rangle$ with longer time than in the ground state $|0,+\rangle$. Finally, for $g_1=0$ in Fig. \ref{Fig3}(c), the oscillation disappears completely and the system is trapped in the target state in an ideal case, which is consistent with the results $\langle P_{0+}\rangle=0$ and $\langle P_{2-}\rangle=1$ in Fig. \ref{Fig2}(b) at $\Delta_{m}\approx 20.6g$.
\begin{figure*}[htbp]
  \centering
	\includegraphics[scale=0.8]{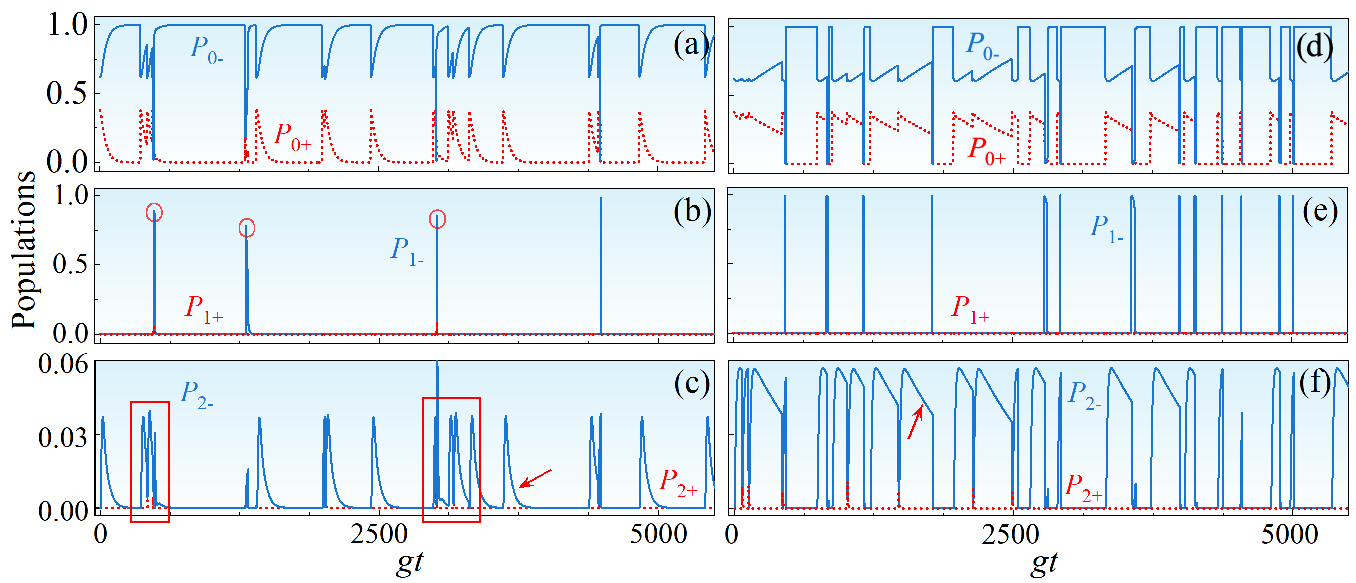}
    \caption{(a)-(c) and (d)-(f) Quantum trajectories of two-magnon bundle emission under conditions of $g_1=g$ and $g_1=0$, respectively. In Fig. (b), the three red circles indicate that the population of the single-magnon state fails to reach unity. In Fig. (c), the two red boxes indicate rapid oscillations between the excited state and the vacuum state on short timescales, while the red arrow represents the decay from the excited state back to the vacuum state without the occurrence of a quantum jump. In Fig. (f), the red arrow indicates that under unidirectional coupling, the excited-state lifetime is prolonged, allowing sufficient time for magnon emission.  
    The system is initiated in the ground state $|0,g\rangle$ and the remaining parameters are identical to those in Fig. \ref{Fig4}.   \label{Fig5}}
\end{figure*}

\section{Enhanced magnon bundle emission} \label{sec3}
In this section, we incorporate dissipation into the system and investigate the full dynamical behavior of magnon bundle emission. By comparing the two cases of unidirectional coupling ($g_1=0$) and symmetric coupling ($g_1=g$), we quantitatively evaluate the influence of full chiral interaction on  bundle purity and emission number.
The statistical properties of the emitted magnons are characterized by the second-order correlation function:
\begin{eqnarray}\label{Eqg2}
    g^{(2)}(\tau)=\frac{\langle m^\dagger(0) m^\dagger(\tau)m(\tau)m(0)\rangle}{\langle(m^\dagger m)(0)\rangle\langle(m^\dagger m)(\tau)\rangle},
\end{eqnarray}
where $\tau$ is the delayed time. The magnons in the bundle behave as a single physical entity, and their time-delayed correlation can be assessed by the above second-order correlation function. For the $N$-magnon bundle, the second-order correlation function is defined as \cite{munoz2014emitters}
\begin{eqnarray}
g^{(2)}_N(\tau)&=&\frac{\langle m^{\dag N}(0) m^{\dag N}(\tau) m^{N}(\tau) m^{N}(0)\rangle}{\langle (m^{\dag N} m ^N)(0)\rangle \langle (m^{\dag N} m ^N)(\tau)\rangle},
\end{eqnarray}
where $\tau$ is the delayed time and $\tau\geq\tau_{\textrm{min}}=\sum_{\ell=1}^{N}\frac{1}{\ell\kappa_m}$. $\tau_\textrm{min}$ can be approximately regarded as zero-time delay for the bundle emission, since in the small-time delay windows of width $\tau_\textrm{min}$ centered on zero, the bundle correlation function is ill-defined (as such short times probe inside the bundle itself) \cite{munoz2014emitters,bin2021parity}.
When $N=1$, $g^{(2)}_N(\tau)$ becomes the standard form defined by Eq. (\ref{Eqg2}). For $N=2$, it describes the second-order correlation function of two-magnon bundle. The magnons in the same bundle are bunching for $g^{(2)}(0)>1$ while the neighboring bundles are antibunched when $g^{(2)}_2(0)<1$. 

In Fig. \ref{Fig4}(a) we plot the equal-time correlation functions $g^{(2)}(0)$ and $g^{(2)}_2(0)$ as a function of $\Delta_{m}/g$. As shown, the dotted curves from the non-Hermitian Hamiltonian align closely with the solid curves from the full master equation, confirming that the two approaches yield equivalent dynamics. The curves exhibit pronounced resonant features (peaks or dips) at  $\Delta_{m}\approx\tilde{\Omega}/N$. In particular, at $\Delta_{m}=20.6g$, we find that $g^{(2)}(0)>1$ and $g^{(2)}_2(0)<1$ are simultaneously satisfied. Furthermore, in Figs. \ref{Fig4}(b) and  \ref{Fig4}(c), we plot the delayed-time second-order correlation function for two cases under the condition of two-magnon resonance. On the one hand, in both cases we find $g^{(2)}(\tau)<g^{(2)}(0)$ and $g^{(2)}_2(\tau)>g^{(2)}_2(0)$ in the effective regions $\tau>\tau_\textrm{min}$, indicating that the emitted magnons in a bundle are strongly bunched, whereas the bundles themselves are antibunching.
On the other hand, compared with the symmetric coupling scheme, the unidirectional case leads to a significant enhancement of $g^{(2)}(\tau)$ and a strong suppression of $g^{(2)}_2(\tau)$. This indicates that both the bunching effect of magnons within a bundle and the antibunching effect between bundles are strengthened.
Obviously, the quality of the emitted bundle would be improved in the one-way coupling case, i.e., $g_{1}=0$, which is termed as chiral enhanced bundle emission.

To describe the entire bundle emission process, we employ Monte Carlo simulations to record the individual quantum trajectories of the system. For the unidirectional coupling case, the simulations are based on the non-Hermitian Hamiltonian, whose dynamics are equivalent to those of the full master equation, as demonstrated in Fig. \ref{Fig4}(a). 
Here, we present two representative quantum trajectories for $g_1=0$ and $g_1=g$, as shown in Fig. \ref{Fig5}, to clearly reveal the differences in the dynamical evolution. In a complete two-magnon bundle emission cycle, the system is initially prepared in the ground state $|0,g\rangle$ and is driven into the target state $|2,-\rangle$. Subsequently, a single magnon is triggered by the dissipation and the other correlated one follows quickly through the intermediate state $|1,-\rangle$. The system finally returns to the initial state from $|0,-\rangle$ to $|0,+\rangle$ through the atomic dissipation forming a bundle emission cycle, which is presented in Fig. \ref{Fig5}. By comparing the dynamical evolution of the chiral interaction with the achiral case, there are several advantages as follows.
In the first place, as shown in Figs. \ref{Fig5}(c) and \ref{Fig5}(f), the excitation probability $P_{2-}$ is substantially enhanced from 0.04 to 0.06, indicating that the chiral interaction facilitates the preparation of the high Fock state. Secondly, we see in Fig. \ref{Fig5}(e) that the maximal probability of $P_{1-}$ always approaches unity, which is hard to realize for the symmetrical case as indicated by red circles in Fig. \ref{Fig5}(b).  This behavior arises because the system can radiate a magnon from a non‑target state with a finite probability, such as $|1,-\rangle$. It implies that the purity of magnon bundle is possible to boost in the one-way coupling scheme. Finally, as shown in Fig. \ref{Fig5}(f), the magnon stays in the target excited state with a significantly longer duration than that in Fig. \ref{Fig5}(c). The prolonged excited-state lifetime arises because, for $g_1=0$, the reverse transition channel $|2,-\rangle\to|0,+\rangle$, governed by the term $m^2\sigma_{+-}$ in Eq. (\ref{Eq10}), is closed. In the absence of dissipation, the closure of this transition channel turns the system from the so-called super-Rabi oscillation [Fig. \ref{Fig3}(a)] to eventual trapping in the target excited state [Fig. \ref{Fig3}(c)]. Consequently, even when the dissipation of the magnon and the atom is taken into account, the blocked transition channel ensures that the system prefers to stay in the target excited state. 

\begin{figure}[htbp]
  \centering
	\includegraphics[scale=0.6]{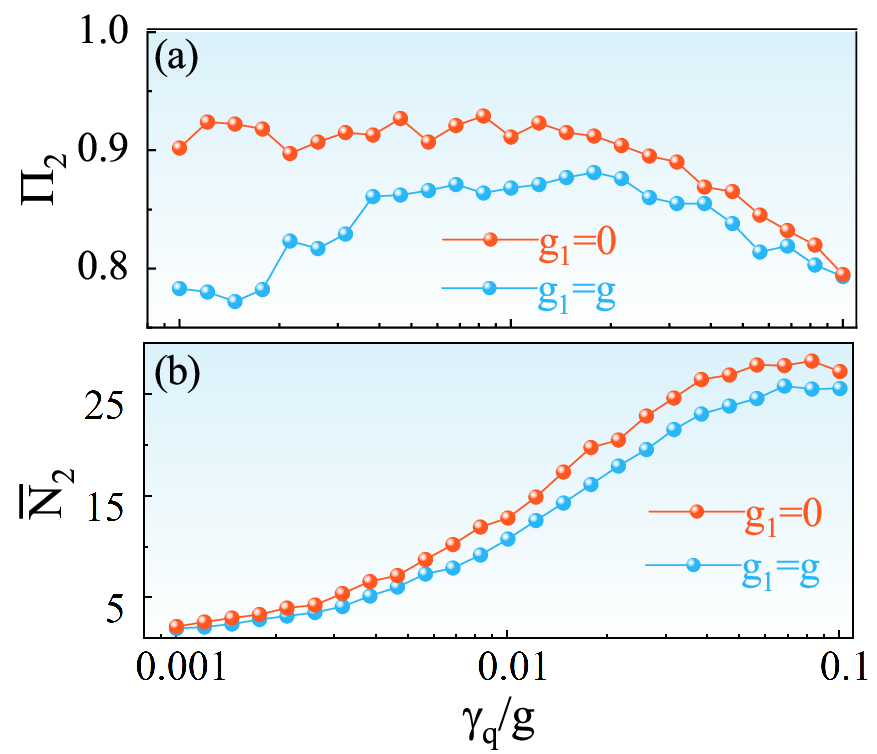}
    \caption{The average purity and number of bundles as a function of the atomic dissipation rate $\gamma_q$. In addition, the time duration for each trajectory is set to $\Delta T = 10^4/g$ and each data point is derived from a statistical average over 100 trajectories.  The other parameters are the same as in Fig. \ref{Fig5}. \label{Fig6}}
\end{figure}

Furthermore, the internal mechanisms of the chirality-enhanced magnon bundle emission are analyzed as follows. On one hand, as indicated by the red arrow in Fig. \ref{Fig5}(c), the prepared excited Fock state $|2,-\rangle$ decays slowly back to the ground state without undergoing quantum jumps. Such behavior occurs repeatedly thus reducing the utilization of excitation cycles. On the other hand, the situation is different in the case of $g_1=0$, wherein the lifetime of the excited state is significantly prolonged, allowing sufficient time for quantum jumps to occur as indicated by the red arrow in Fig. \ref{Fig5}(f). 
More importantly, the re-excitation of the magnon mode is reasonably suppressed since the excited state is occupied with the extended lifetime. However, in the normal symmetrical case, the excited state quickly turns back to the lower state several times in a short period (highlighted by the red boxes in Fig. \ref{Fig5}(c)), and this behavior greatly increases the probability of the system being driven into non‑target states, preventing the probability of state $|1,-\rangle$ in Fig. \ref{Fig5}(b) from reaching unity. As a result, the emission number and the purity of bundles would be improved simultaneously based on the chirality-dependent interaction. To verify this point, we numerically simulated the bundle emission event by using the same parameters in Fig.  \ref{Fig5}. It is clear that the average purity of bundles rises from 86.8\% for symmetric coupling to 91.1\% for  chiral coupling and the average number is increased from 10.75 to 12.82.

To clearly demonstrate the influence of chiral interaction on the average number and purity of magnon bundles, we plot the average number $\bar{N}_2$ and purity $\bar{\Pi}_2$ versus atomic dissipation rate $\gamma_q$, as shown in Fig. \ref{Fig6}. The purity of bundle emission is defined as $\Pi_N=\bar{N}_N/\sum\bar{N}_j$, where $\bar{N}_j$ represents the statistical average number of $j$-magnon emissions. As shown in Fig. \ref{Fig6}(a), the purity of magnon bundles is always enhanced in the asymmetric case. Notably, when the atomic decay rate $\gamma_{q}$ is low, the increase in purity is much greater than in the case where the larger values of $\gamma_{q}$ are chosen. Physically, the smaller the atomic decay rate is, the longer the time interval between adjacent bundles will be, leading to higher purity of two-magnon bundle emission. In particular, by choosing the optimal parameter region (around $\gamma_q=0.01g$) of bundle emission, the increase in purity is higher than 5\%. In addition, as shown in Fig. \ref{Fig6}(b), with the increase of atomic decay rate, the average number of bundle emission for the achiral case also rises slowly. These findings are in good agreement with the theoretical prediction above.

\begin{figure}[htbp]
  \centering
	\includegraphics[scale=0.62]{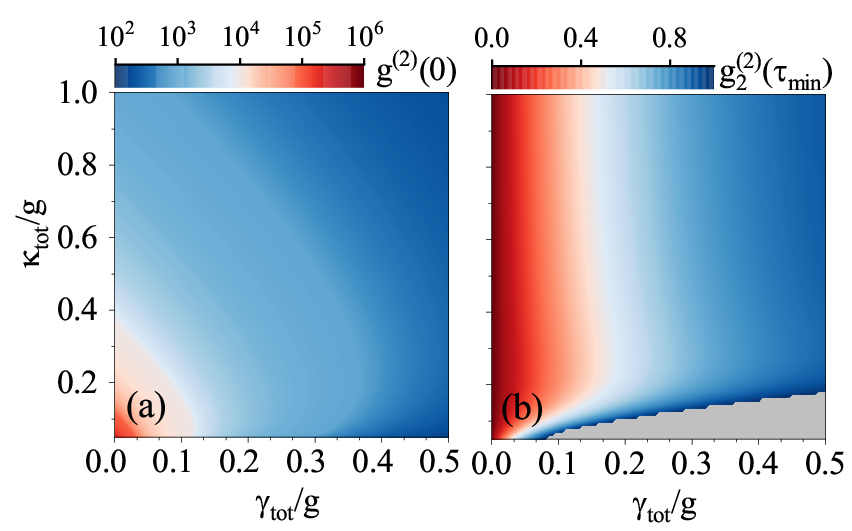}
    \caption{The correlation function (a) $g^{(2)}(0)$ and (b) $g_2^{(2)}(\tau_{min})$ as a function of $\kappa_\textrm{tot}$ and $\gamma_\textrm{tot}$. The shaded region in (b) indicates $g_2^{(2)}(\tau_\textrm{min})>1$. The other parameters are the same as in Fig. \ref{Fig6}. \label{Fig7}}
\end{figure}

In the preceding analysis, the cavity photon loss rate is assumed to be low enough to guarantee that the cavity-mediated dissipation of the magnon and the atom was far smaller than the environmental dissipation. Here, we show the combined effects of total dissipation $\kappa_\textrm{tot}$ and $\gamma_\textrm{tot}$ on the correlated properties of the magnon bundle, in which both the environmental and cavity‑loss‑induced contributions are considered. 
As shown in Fig. \ref{Fig7}(a), $g^{(2)}(0) >1$ holds across the entire parameter range, indicating that the magnons are bunching. In Fig. \ref{Fig7}(b), apart from the shaded region, the other areas satisfy $g_2^{(2)}(\tau_\textrm{min})<1$, meaning that the bundles are antibunching. These two figures demonstrate that magnon bundles emission can be realized over a broad dissipation regime. However, it should be emphasized that while bundles can still be generated, their quality is significantly degraded in the strong‑dissipation regime.

Before ending this section, we outline the potential experimental implementations of the proposed system. As shown in Fig. \ref{Fig1}(a), here we adopt a cascaded configuration of two microwave cavities, into which a qubit and a YIG sphere are placed, respectively. To realize near-ideal unidirectional coupling, the output ports of the two cavities are connected through a circulator rather than coupled directly to the transmission line. 
Experimentally, strong coupling between a superconducting qubit and a microwave cavity has been demonstrated in circuit-QED platforms \cite{Schuster2004cQED, Hofheinz2009Synthesizing,Christensen2019AnomalousChargeNoise,Place2021TantalumTransmon}.  For example, a superconducting qubit coupled to a resonator was reported to have a coupling strength of $g/2\pi = 100~\mathrm{MHz}$ \cite{Christensen2019AnomalousChargeNoise}. Moreover, state-of-the-art superconducting transmon qubits have exhibited lifetimes and coherence times exceeding $0.3~\mathrm{ms}$, corresponding to decoherence rates in the kHz range, far below the MHz scale \cite{Place2021TantalumTransmon}.
On the other hand, an undoped single-crystal YIG sphere is installed in the right cavity. As mentioned in Ref. \cite{zhang2014strongly,tabuchi2014hybridizing,bourhill2016ultrahigh,kostylev2016superstrong,yuan2022quantum,huebl2013high}, the ultra-strong coupling between magnon and microwave photons has been achieved. According to various experimental reports, the microwave photon–magnon coupling strength can range from tens of MHz to 2.5 GHz, while the magnon dissipation is typically on the order of MHz. Based on these experimental parameters, we adopt the following values for our system: $g/2\pi=10$ MHz, $\kappa_m/2\pi=1$ MHz, $\gamma_q/2\pi=0.1$ MHz, $\Omega/2\pi=200$ MHz, $\Delta_q/2\pi=100$ MHz and $\Delta_m/2\pi=206$ MHz. 
It is reported in experiment that the magnon emission in YIG can be detected via state tomography, which allows reconstruction of the density matrix of the magnetization dynamical states \cite{Hioki2021State,Xu2023Quantum}. Alternatively, magnon states can be read out by coupling the magnon mode to an auxiliary superconducting qubit operating in the strong dispersive regime \cite{Lachance2017Resolving,Lachance2020Entanglement}.

\section{Conclusion} \label{sec4}
In summary, we have demonstrated that the fully chiral coupling between a qubit and a magnon mode is realized through a cascaded cavity setup, which provides  an internal directional interaction responsible for the enhancement of two-magnon bundle emission. By eliminating the conjugate term of the magnon-qubit interaction, the system is prepared into the target state with longer lifetime and then the re-excitation of the magnon mode to the higher Fock state is obviously suppressed, leading to an increase in the average purity and number of magnon bundles. It opens an effective avenue to prepare directional magnon sources with high quality and may find potential applications in quantum information processing and quantum sensing technology. 

\begin{appendix}

\section{The Effective Hamiltonian}\label{secA}
In the dressed picture, the system Hamiltonian Eq. (\ref{EqH_t}) can be rewritten as $H=H_0+V$:
\begin{eqnarray}
\nonumber
H_0&=&\Delta_m m^{\dag}m +\lambda_+\sigma_{++}+\lambda_-\sigma_{--},\\  \nonumber
V&=&\eta_{-+}(g_1m\sigma_{+-}+g_2m^{\dag} \sigma_{-+}) \\ \nonumber
&-&\eta_{--}(g_1m\sigma_{--} +g_2m^{\dag} \sigma_{--}) \\  \nonumber
&-&\eta_{+-}(g_1m\sigma_{-+}+ g_2m^{\dag} \sigma_{+-})  \\ 
&+&\eta_{++}(g_1m\sigma_{++}+ g_2m^{\dag} \sigma_{++}),
\end{eqnarray}

where  $\eta_{-+}=\cos^2\theta$, $\eta_{--}=\cos\theta\sin\theta$, $\eta_{+-}=\sin^2\theta$, $\eta_{++}=\cos\theta\sin\theta$. Using the transformation $H_{\textrm{eff}}=e^{-\zeta X}He^{\zeta X}$, the effective Hamiltonian is obtained. The $\zeta$ is introduced to show the orders in the perturbation expansion and would be set to 1 after the calculations. The non-Hermitian operator $X$ is introduced with the form
\begin{eqnarray}
\nonumber
X&=&\frac{\eta_{-+}}{\Delta_m-\tilde{\Omega}}(g_1m\sigma_{+-}-g_2m^\dagger\sigma_{-+}) \\ \nonumber
&-&\frac{\eta_{--}}{\Delta_m }(g_1m\sigma_{--}-g_2m^\dagger\sigma_{--}) \\ \nonumber
&+&\frac{\eta_{++}}{\Delta_m }(g_1m\sigma_{++}-g_2m^\dagger\sigma_{++}) \\ 
&-&\frac{\eta_{+-}}{\Delta_m+\tilde{\Omega}}(g_1m\sigma_{-+}-g_2m^\dagger\sigma_{+-}),
\end{eqnarray}
which is satisfied as $[H_{\textrm{0}},X]=-V$. In terms of the Baker-Campbell-Hausdorff formula, one has
\begin{eqnarray}
\nonumber
H_{\textrm{eff}}^1 &=& H_0+\zeta\left(V+[H_0,X]\right) \\
\nonumber
&+&\zeta^2\left([V,X]+\frac{1}{2}[[H_0,X],X]\right)+O(\zeta^3) \\
&=& H_0+\frac{\zeta^2}{2}[V,X]+O(\zeta^3),
\end{eqnarray}
Under the two-magnon resonance condition $\Delta_m=\tilde{\Omega}/2$, the effective Hamiltonian is obtained as 
\begin{eqnarray}\label{A4}
H_{\textrm{eff}}^1=g_{\textrm{eff}}^1m^2\sigma_{+-}+g_{\textrm{eff}}^2m^{\dagger 2}\sigma_{-+},
\end{eqnarray}
where $g_{\textrm{eff}}^1=-4g^2_1\eta_{-+}\eta_{++}/\tilde{\Omega}$, $g_{\textrm{eff}}^2=-4g^2_2\eta_{-+}\eta_{++}/\tilde{\Omega}$. Similarly, by choosing $\Delta_m=-\tilde{\Omega}/2$, we obtain another transition channel and the  corresponding Hamiltonian is
\begin{eqnarray}\label{A5}
H_{\textrm{eff}}^1=\tilde{g}_{\textrm{eff}}^1m^2\sigma_{-+}+\tilde{g}_{\textrm{eff}}^2m^{\dagger 2}\sigma_{+-},
\end{eqnarray}
where $\tilde{g}_{\textrm{eff}}^1=4g^2_1\eta_{+-}\eta_{++}/\tilde{\Omega}$, $\tilde{g}_{\textrm{eff}}^2=4g^2_2\eta_{+-}\eta_{++}/\tilde{\Omega}$.

\end{appendix}

\section*{Acknowledgments}
This work is supported by the National Natural Science Foundation of China (Grants Nos. 12274164, 61875067, 12375011 and 12304392).\\

\end{document}